\documentclass[pra,twocolumn,showpacs,preprintnumbers,amsmath,amssymb]{revtex4}
\usepackage{amsfonts}
\usepackage{tipa}

\usepackage{epsfig,graphicx}
\usepackage{amstext}
\usepackage{amsmath}
\usepackage{graphicx}

\begin{document}


\title{Robustness of quantum correlations against decoherence}

\author{Ming-Liang Hu$^{1}$}
\author{Heng Fan$^{2}$}
\address{$^{1}$School of Science, Xi'an University of Posts and
               Telecommunications, Xi'an 710061, China \\
         $^{2}$Beijing National Laboratory for Condensed Matter Physics,
               Institute of Physics, Chinese Academy of Sciences, Beijing
               100190, China}

\begin{abstract}
We study dynamics of nonclassical correlations by exactly solving a
model consisting of two atomic qubits with spontaneous emission. We
find that the nonclassical correlations defined by different
measures give different qualitative characterizations of those
correlations. The relative behaviors of those correlation measures
are presented explicitly for various quantum states in the two-qubit
atomic system. In particular, we find that the robustness of quantum
correlations can be greatly enhanced by performing appropriate local
unitary operations on the initial state of the system.
\end{abstract}

\pacs{03.65.Ta, 03.67.-a, 03.65.Yz \\Key Words: Entanglement;
Measurement-induced disturbance; Quantum discord; Geometric measure
of quantum discord}

\maketitle

\section{Introduction}
Entanglement plays a key role and is a resource in quantum
information processing (QIP). In the last decades, much efforts have
been put into study it in various physical systems, see for example
Refs. \cite{Bennett,Wootters,Horodecki,Yu,Amico}. While being widely
considered to be something unique without classical counterpart, it
is recently realized that entanglement reflects only one aspect of
quantum correlation of a quantum state, there exist some other
nonclassical correlations \cite{Ollivier,Luo,Dakic,Modi,Streltsov}.
One example is a model of QIP in Ref. \cite{knill}, which
outperforms its classical counterpart while without any
entanglement. It is proposed that quantum discord (QD)
\cite{Ollivier}, a different measure of quantum correlation, might
be responsible for the power of this model \cite{Datta}. Thus,
entanglement and QD, and additionally some other measures all are
important in describing nonclassical correlations for a quantum
system.

It is well acceptable that decoherence will cause decreasing of
entanglement which may induce failure of the algorithms and various
protocols of QIP. On the other hand, it is also of fundamental
interest to study the behavior of nonclassical correlations with
decoherence. Recently, there are some studies in evaluating the
relationships between nonclassical correlations quantified by QD and
entanglement quantified by concurrence or entanglement of formation
(EoF) under the action of decoherence
\cite{Werlang,Wang,Fanchini,Maziero,Luxm,Liy,Ferraro,Fus,Luos,Ali,Lang,Mazzola}.
Also the role of QD in identifying critical points of quantum phase
transitions are studied \cite{Trippe,Sarandy,Liu,Liyc}. For
practical use, one may hope that the nonclassical correlations which
are crucial for QIP can be maintained for sufficiently long time so
that the designed tasks can be fulfilled. However, the unavoidable
interaction of a realistic system with its surroundings always makes
it decaying with time. Particularly, under certain circumstances,
the entanglement of a bipartite state can even terminate abruptly in
a finite time, a phenomenon termed as entanglement sudden death
(ESD) by Yu and Eberly \cite{Yu}. For nonclaccical correlations as
measured by QD and other similar quantities, however, the sudden
death phenomenon has not been observed (this may be good for QIP
independent of entanglement), but their values may decay with time
and become zero at discrete instants of time \cite{Wang,Fanchini}.
Thus it is of great importance to find ways to preserve them when
the system is in contact with an environment which leads to
decoherence.

In this paper, we offer a comparative study of the relationships
between entanglement and different measures of nonclassical
correlations for a certain family of two-qubit states. The system we
considered consists of two atomic qubits with spontaneous emission
\cite{Ficek1}. We will show that even in cases where entanglement
disappears, nonclassical correlations persist in the whole time
region. In this sense, they are more robust against decoherence than
entanglement, implying that quantum algorithms based only on them
may be more robust than those based on entanglement. Moreover, we
will show by explicit examples that robustness of quantum
correlations can be greatly enhanced by performing local unitary
operations to the initial states, though it is well-known that those
measures of correlations themselves are invariant under local
unitary operations \cite{Ollivier,Luo,Dakic}.

This paper is organized as follows. In Section II we recall some
measures of correlations for a bipartite state. We will adopt the
EoF as the measure of entanglement, and other nonclassical
correlations will be quantified by the measurement induced
disturbance (MID), QD and the geometric measure of QD (GMQD). In
Section III, we introduce the model and compare characteristics of
various correlation measures for a class of two-qubit states. We
will also evaluate robustness of quantum correlations for different
local unitary equivalent states. Finally, Section IV is devoted to a
summary.

\section{Measures of quantum correlations}
The basic and crucial problem for studying dynamics of correlations
of a composite system is to characterize and quantify correlation
from different point of views in that system. The motivation of this
study is to find whether those physical systems can be a platform
for various tasks of QIP. In this section, we will first list
several measures of quantum correlations for the bipartite states,
and those measures will be applied in our study of the two-qubit
atomic system.

First, we still consider entanglement as one important quantum
correlation. In order to study the entanglement dynamics for any
two-qubit states, we adopt EoF \cite{Bennett} which is a
well-accepted measure of entanglement. The definition is as follows,
\begin{equation}
 E=H\left(\frac{1+\sqrt{1-C^2}}{2}\right),
\end{equation}
where $H(\tau)=-\tau\log_{2}\tau-(1-\tau)\log_{2}(1-\tau)$ is the
binary Shannon entropy, and $C$ is the time-dependent concurrence
\cite{Wootters}, which takes the form
$C=\max\{0,\lambda_1-\lambda_2-\lambda_3-\lambda_4\}$, where
$\lambda_i$ ($i=1,2,3,4$) are the square roots of the eigenvalues of
the spin-flipped operator
$R=\rho(\sigma^2\otimes\sigma^2)\rho^*(\sigma^2\otimes\sigma^2)$
arranged in decreasing order, with $\rho^*$ being the complex
conjugation of $\rho$ in the standard basis, and $\sigma^2$ is one
Pauli matrix.

Besides entanglement, let us consider some other quantum
correlations. The so-called MID, proposed recently by Luo in Ref.
\cite{Luo}, is defined as the difference of two quantum mutual
information respectively of a given state $\rho$ shared by two
parties $a$ and $b$ and the corresponding post-measurement state
$\Pi(\rho)$,
\begin{equation}
 \text{MID}(\rho)=I(\rho)-I(\Pi(\rho)),
\end{equation}
where the mutual information, $I(\rho)=S(\rho^a)+S(\rho^b)-S(\rho)$,
measures the total correlation, including both classical and
quantum, for a bipartite state $\rho$. Here
$S(\rho)=-\text{tr}(\rho\log_2\rho)$ denotes the von Neumann
entropy, with $\rho^a$ and $\rho^b$ being the reduced density matrix
of $\rho$ by tracing out $b$ and $a$, respectively. While quantum
mutual information of the state $\Pi(\rho)$, $I(\Pi(\rho))$,
quantifies the classical correlation in $\rho$, with
$\Pi(\rho)=\Sigma_{ij}(\Pi_i^a\otimes\Pi_j^b)\rho(\Pi_i^a\otimes\Pi_j^b)$,
where the measurement is induced by the spectral resolutions of the
reduced states $\rho^a=\Sigma_i p_i^a\Pi_i^a$ and $\rho^b=\Sigma_j
p_j^b\Pi_j^b$. In fact, $\Pi(\rho)$ is a classical state (a state is
referred to as classical if it is invariant under local von Neumann
measurement) for any complete set of projective measurements
$\{\Pi_i^a\}$ and $\{\Pi_j^b\}$. $\Pi(\rho)$ defined above is
actually the closest classical state to $\rho$ since this kind of
measurement leaves the marginal information invariant and is in a
certain sense the least disturbing \cite{Luo}.

Another well-accepted measure of nonclassical correlation is the QD
\cite{Ollivier}. It was defined by the discrepancy between quantum
mutual information and the classical aspect of correlation which was
defined as the maximum information about one subsystem that can be
obtained by performing a measurement on the other subsystem. If we
restrict ourselves to the projective measurements performed locally
on subsystem $a$ described by a complete set of orthogonal
projectors $\{\Pi_k\}$, then the quantum state after measurement
changes to
$\rho_{b|k}=(\Pi_k\otimes\mathbb{I})\rho(\Pi_k\otimes\mathbb{I})/p_k$,
where $\mathbb{I}$ is the identity operator for subsystem $b$, and
$p_k=\text{tr}[(\Pi_k\otimes\mathbb{I})\rho(\Pi_k\otimes\mathbb{I})]$
is the probability for obtaining the measurement outcome $k$ on $a$.
The classical correlation can then be obtained by maximizing
$J(\rho|\{\Pi_k\})=S(\rho^b)-S(\rho|\{\Pi_k\})$ over all
$\{\Pi_k\}$, where $S(\rho|\{\Pi_k\})=\Sigma_k p_k S(\rho_{b|k})$ is
a generalization of the classical conditional entropy of subsystem
$b$. Explicitly, the QD is defined as the minimal difference between
$I(\rho)$ and $J(\rho|\{\Pi_k\})$ as
\begin{equation}
 \text{QD}(\rho)=I(\rho)-\sup_{\{\Pi_k\}}J(\rho|\{\Pi_k\}),
\end{equation}
where the supremum is taken over the complete set of $\{\Pi_k\}$.
The intuitive meaning of QD thus may be interpreted as the minimal
loss of correlations due to measurement. It vanishes for states with
only classical correlation and survives for states with quantum
correlation.

Since analytical expressions of QD are achievable only for certain
special classes of states \cite{Luos,Ali}, Daki\'{c} {\it et al.}
\cite{Dakic} proposed the GMQD. They use the square of the
Hilbert-Schmidt norm as the distance between two quantum states, and
the nearest distance between $\rho$ and all of the zero-discord
states $\Omega_0$ is interpreted as a measure of quantum
correlation, which is defined as

\begin{equation}
 \text{GMQD}(\rho)=\min_{\chi\in\Omega_0}\parallel\rho-\chi\parallel^2,
\end{equation}
where the geometric quantity
$\parallel\rho-\chi\parallel^2=\text{tr}(\rho-\chi)^2$.

By noting that any two-quit state $\rho$ can be represented as
$\rho=\frac{1}{4}(\mathbb{I}\otimes\mathbb{I}+\vec{x}\cdot\vec{\sigma}\otimes\mathbb{I}+
 \mathbb{I}\otimes\vec{y}\cdot\vec{\sigma}+\sum_{i,j=1}^3r_{ij}\sigma_i\otimes\sigma_j)$,
Daki\'{c} {\it et al.} derived an explicit formula of GMQD, which is
given by \cite{Dakic}
\begin{equation}
 \text{GMQD}(\rho)=\frac{1}{4}\left(\parallel x\parallel^2+\parallel R\parallel^2-k_\text{max}\right),
\end{equation}
where the scalar product
$\vec{\alpha}\cdot\vec{\sigma}=\sum_{i=1}^3\alpha_i\sigma_i$ with
$\alpha=x,y$. Moreover, $||x||^2=\sum_{i=1}^3 x_i^2$,
$||R||^2=\text{tr}(R^T R)$, and $k_\text{max}$ is the largest
eigenvalue of the matrix $K=xx^T+RR^T$, where the superscript $T$
denotes transpose of vectors or matrices. Since this measure gives
analytic results without restriction on the form of $\rho$, it turns
out to be a convenient tool for analyzing quantum correlation
dynamics from a geometric perspective.

Correlation measures (i.e.,EoF, MID and QD) listed above assume
equal values for bipartite pure states, but this is not the case for
mixed states. Particularly, the QD and GMQD are not symmetric
quantities with respect to the measurements performed on subsystem
$a$ or $b$, and the state is said to be completely classically
correlated only when both of them reach zero. In the subsequent
discussions, however, we will restrict ourselves to the situation
that the states are symmetric under exchange of subsystems.

\section{Correlation dynamics under decoherence}
We are interested in revealing difference between various
correlation measures under the action of noisy environments. For
this purpose, we consider a system consists of two identical atoms
with spontaneous emission, and having lower and upper levels
$|g_i\rangle$ and $|e_i\rangle$ ($i=1,2$) separated by the energy
gap $\hbar\omega$, with $\omega$ being the transition frequency.
Also, we assume this system is coupled to a multimode vacuum
electromagnetic field, under the influence of which its time
evolution is governed by the following master equation
\cite{Ficek1,Ficek2}
\begin{eqnarray}
 \frac{\partial\rho}{\partial t}&=&-\text{i}\omega\sum_{i=1}^2
 [S_i^z,\rho]-\text{i}\sum_{i\neq j}^2\Omega_{ij}[S_i^+S_j^-,\rho]\nonumber\\&&
 +\frac{1}{2}\sum_{i,j=1}^2\gamma_{ij}\left(2S_j^{-}\rho S_i^{+}
 -\{S_i^{+}S_j^{-},\rho\}\right),
\end{eqnarray}
where $S_i^\pm$ are the raising and lowering operators while $S_i^z$
is the energy operator. $\gamma_{ij}\equiv\gamma$ ($i=j$) are the
spontaneous emission rates of the atoms caused by their direct
interaction with the vacuum field. Moreover, $\gamma_{ij}$ and
$\Omega_{ij}$ ($i\neq j$) describe the collective damping and the
dipole-dipole interaction potential, respectively. They both depend
on the interatomic distance $r_{ij}=|r_j-r_i|$ and take the form
\begin{eqnarray}
 \gamma_{ij}&=&\frac{3}{2}\gamma\left[\frac{\sin(kr_{ij})}{kr_{ij}}+\frac{\cos(kr_{ij})}{(kr_{ij})^2}
 -\frac{\sin(kr_{ij})}{(kr_{ij})^3}\right], \nonumber\\
 \Omega_{ij}&=&\frac{3}{4}\gamma\left[\frac{\sin(kr_{ij})}{(kr_{ij})^2}+\frac{\cos(kr_{ij})}{(kr_{ij})^3}
 -\frac{\cos(kr_{ij})}{kr_{ij}}\right],
\end{eqnarray}
where we have assumed that the atomic dipole moments for the two
atoms take the same value (i.e., $\mu_1=\mu_2=\mu$) and are
polarized in the direction perpendicular to the interatomic axis.
And, $k=2\pi/\lambda$ is the wave vector with $\lambda$ being the
atomic resonant wavelength.

We first consider the Bell-like initial state as follows
\begin{equation}
 |\Psi\rangle=\alpha|e_1 e_2\rangle+\sqrt{1-\alpha^2}|g_1 g_2\rangle,
\end{equation}
for which the nonzero elements of $\rho(t)$ in the standard basis
$\{|e_1 e_2\rangle,|e_1 g_2\rangle,|g_1 e_2\rangle,|g_1
g_2\rangle\}$ are of the form
\begin{eqnarray}
 \rho_{11}(t)&=&\alpha^2 e^{-2\gamma t},\nonumber\\
 \rho_{14}(t)&=&\rho_{41}^*(t)=\alpha\sqrt{1-\alpha^2} e^{-(\gamma+2\text{i}\omega)t},\nonumber\\
 \rho_{22,33}(t)&=&a_1[e^{-\gamma_{12}^{+}t}-e^{-2\gamma t}]
                   +a_2[e^{-\gamma_{12}^{-}t}-e^{-2\gamma t}],\nonumber\\
 \rho_{23,32}(t)&=&a_1[e^{-\gamma_{12}^{+}t}-e^{-2\gamma t}]
                   -a_2[e^{-\gamma_{12}^{-}t}-e^{-2\gamma t}],\nonumber\\
 \rho_{44}(t)&=&1-\rho_{11}(t)-\rho_{22}(t)-\rho_{33}(t),
\end{eqnarray}
where $\gamma_{12}^{\pm}=\gamma\pm\gamma_{12}$, $a_{1,2}=\alpha^2
\gamma_{12}^{\pm}/2\gamma_{12}^{\mp}$. As one can see, the X
structure of $\rho(t)$ is maintained during the evolution, which
greatly facilitates the following analysis.
\begin{figure}
\centering
\resizebox{0.45\textwidth}{!}{%
\includegraphics{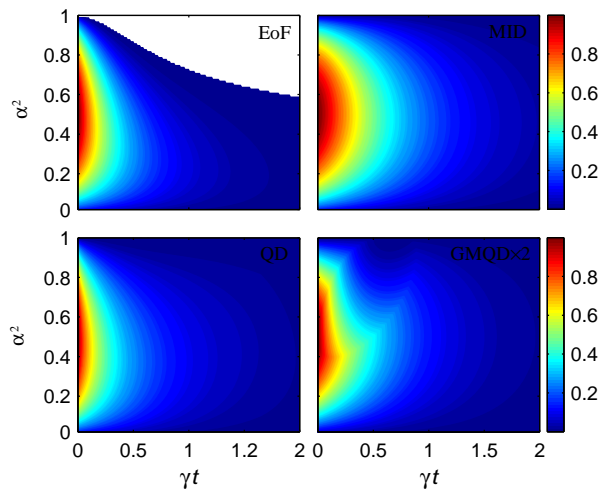}}
\caption{(Color online) Contour plots of different correlation
measures versus $\gamma t$ and $\alpha^2$ for the initial state
$|\Psi\rangle$, where the interatomic distance for these plots is
$r_{12}=0.6737\lambda$.} \label{fig:1}
\end{figure}

Fig. 1 shows the contour plots of the quantum correlations versus
$\gamma t$ and $\alpha^2$ in the short time region with the
interatomic distance $r_{12}=0.6737\lambda$, at which $\gamma_{12}$
reaches its minimal value. Clearly, while the ESD happens after a
finite time, the MID, QD and GMQD maintain during the whole time
region. This points to a fact that the nonclassical correlations may
be more resistant to external perturbations than that of
entanglement. Moreover, the GMQD shows a revival for large values of
$\alpha^2$, but during the same time region the other correlation
measures are always decreased. Thus similar to the relativity of
different entanglement measures (see \cite{Miranowica} and
references therein), the nonclassical correlation measures may also
impose different orderings of quantum states.
\begin{figure}
\centering
\resizebox{0.4\textwidth}{!}{%
\includegraphics{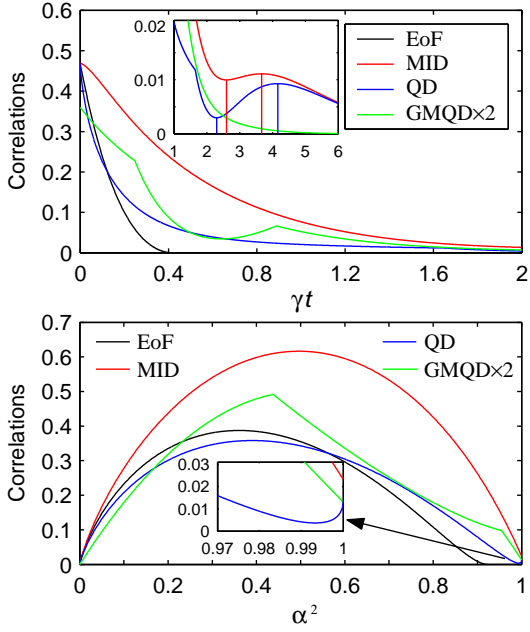}}
\caption{(Color online) Quantum correlations versus $\gamma t$ with
$\alpha^2=0.9$ (top) and versus $\alpha^2$ with $\gamma t=0.35$
(bottom) for the initial state $|\Psi\rangle$ with
$r_{12}=0.6737\lambda$. The insets show the cases when $\gamma
t\in[1,6]$ and $\alpha^2\in[0.97,1]$, where the curves are cut to
better visualize of the revival phenomenon, and the vertical lines
indicate the critical $\gamma t$ at which the correlations turn out
to be increased or decreased.} \label{fig:2}
\end{figure}

For quantum states of Eq. (9), the concurrence can be obtained
analytically as $C(\rho)=2\max\{0,C_1,C_2\}$, where
$C_1=|\rho_{14}|-\sqrt{\rho_{22}\rho_{33}}$ and
$C_2=|\rho_{23}|-\sqrt{\rho_{11}\rho_{44}}$. In the short time
region, $|\rho_{14}|^2>\rho_{22}\rho_{33}$, thus the concurrence as
well as EoF are continuous functions of $\gamma t$. In the long time
region, $|\rho_{23}|^2>\rho_{11}\rho_{44}$, and the entanglement may
experiences a very weak revival after a finite time interval of its
complete disappearance. Similarly, for $\rho(t)$ of Eq. (9) we have
$\text{MID}(\rho)=S(\Pi(\rho))-S(\rho)$, with
$\Pi(\rho)=\text{diag}\{\rho_{11},\rho_{22},\rho_{33},\rho_{44}\}$,
thus it is understandable that the MID also behaves as continuous
functions of $\gamma t$ and $\alpha^2$. For the QD and GMQD, as
denoted by the blue and green curves in Fig. 2, they present sudden
changes at discrete instants of time or $\alpha^2$ which are
evidenced by the presence of kinks. In general, these sudden change
behaviors are caused by the optimal procedure for choosing the
measurement operators, and we will discuss this issue in detail in
the following.

From Fig. 2 one can also observe that the MID, QD and GMQD behave as
non-monotonic decreasing functions of $\gamma t$ and $\alpha^2$.
Their values may be increased or decreased during certain parameter
regions. Particularly, it is worthwhile to note that there are
states having different orderings induced by different correlation
measures. For example, for the parameters chosen in Fig. 2 there
exists states $\rho_1$ and $\rho_2$ such that
$\text{MID}(\rho_1)>\text{MID}(\rho_2)$,
$\text{QD}(\rho_1)>\text{QD}(\rho_2)$, while
$\text{GMQD}(\rho_1)<\text{GMQD}(\rho_2)$ when $\gamma
t\in[0.639,0.888]$, i.e., the decrease of MID and QD are accompanied
by the increase of GMQD. Moreover, from the insets of Fig. 2, other
states differently ordered by the nonclassical correlation measures
are readily distinguished, including those during the scaled time
regions $\gamma t\in[2.304,2.604]$ and $\gamma t\in[3.672,4.164]$,
for which we have $\text{MID}(\rho_1)>\text{MID}(\rho_2)$ and
$\text{GMQD}(\rho_1)>\text{GMQD}(\rho_2)$, while
$\text{QD}(\rho_1)<\text{QD}(\rho_2)$. Thus we see that there are no
simple dominance relations between MID, QD and GMQD for general
cases. They are different not only quantitatively but also
qualitatively.

\begin{figure}
\centering
\resizebox{0.4\textwidth}{!}{%
\includegraphics{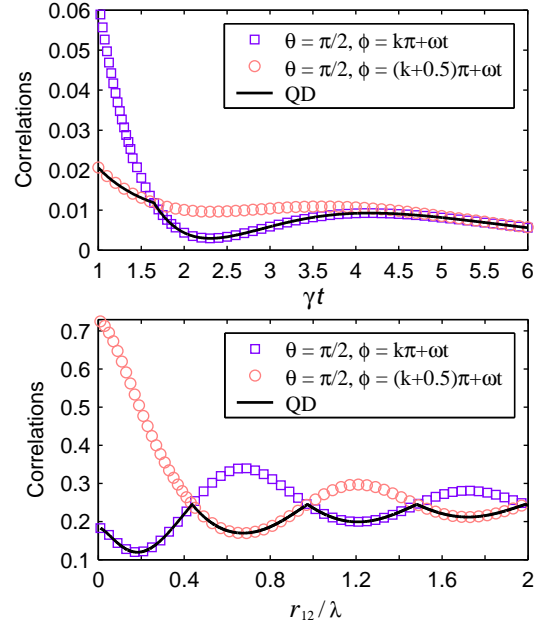}}
\caption{(Color online) QD versus $\gamma t$ with
$r_{12}=0.6737\lambda$, $\alpha^2=0.9$ (top) and versus
$r_{12}/\lambda$ with $\gamma t=0.35$, $\alpha^2=0.8$ (bottom) for
the initial state $|\Psi\rangle$.} \label{fig:3}
\end{figure}

Now we evaluate analytically the sudden change behaviors of the QD.
For $\rho(t)$ in Eq. (9), the infimum of the conditional entropy can
be determined as
\begin{equation}
 \inf_{\{\theta,\phi\}}S(\rho|\{\Pi_k\})=H(\tau),
\end{equation}
where the optimal angle $\theta$ related to the projective
measurement is $\theta=(m+0.5)\pi$ with $m\in\mathbb{Z}$, and
$H(\tau)$ is the Shannon entropy functional, with
\begin{equation}
 \tau=\frac{1-\sqrt{[1-2(\rho_{11}+\rho_{33})]^2+4(|\rho_{14}|^2+|\rho_{23}|^2+\delta)}}{2},
\end{equation}
with $\delta=2\alpha(1-\alpha^2)^{1/2}e^{-\gamma
t}\rho_{23}\cos(2\phi-2\omega t)$. Clearly, the infimum of
$S(\rho|\{\Pi_k\})$ is obtained whenever $\phi=k\pi+\omega t$ if
$\alpha\rho_{23}>0$, and $\phi=(k+0.5)\pi+\omega t$
($k\in\mathbb{Z}$) if $\alpha\rho_{23}<0$. Thus we see that the
sudden change behaviors of QD here is caused by the optimization
procedure for choosing the optimal measurement angle $\phi$ or
equivalently, the optimal projective measurement operators
$\{\Pi_k\}$ because for the two-qubit case $\Pi_k$ can be expressed
as $\Pi_1=(\mathbb{I}+\vec{n}\cdot\vec{\sigma})/2$ and
$\Pi_2=\mathbb{I}-\Pi_1$, with
$\vec{n}=(\sin\theta\cos\phi,\sin\theta\sin\phi,\cos\theta)^T$ being
a unit vector in $\mathbb{R}^3$. Here it should be emphasized that
the angle $\phi$ that minimizes the QD is time dependent, which is
different from the previous results with real elements of the
density matrix \cite{Werlang,Wang,Fanchini,Maziero,Luxm}. As is
shown evidently in the top panel of Fig. 3 with fixed $r_{12}$ and
$\alpha^2$, the optimal measurement angle $\phi$ assumes the value
$\phi=(k+0.5)\pi+\omega t$ in the short time region, while in the
long time region it is $\phi=k\pi+\omega t$. For fixed $\gamma t$
and $\alpha^2$, as exposed in the bottom panel of Fig. 3, the two
different choices of $\phi$ yields the QD alternatively.
\begin{figure}
\centering
\resizebox{0.4\textwidth}{!}{%
\includegraphics{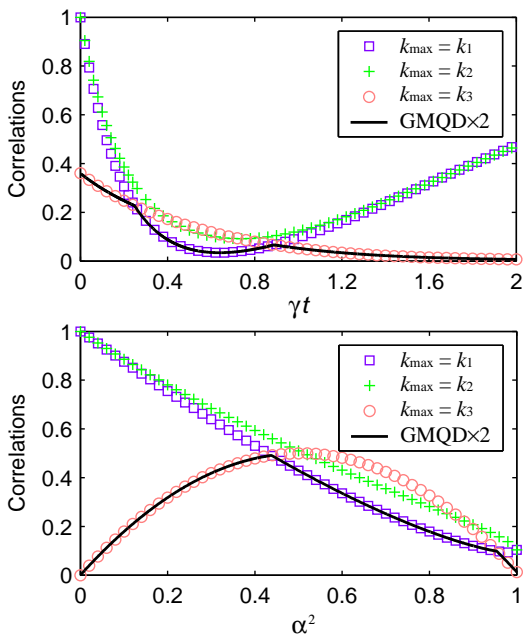}}
\caption{(Color online) GMQD versus $\gamma t$ with $\alpha^2=0.9$
(top) and versus $\alpha^2$ with $\gamma t=0.35$ (bottom) for the
initial state $|\Psi\rangle$ with the interatomic distance
$r_{12}=0.6737\lambda$.} \label{fig:4}
\end{figure}

When considering the GMQD, from Eq. (5) it is readily to see that it
can also be expressed as
$\text{GMQD}(\rho)=(\text{tr}K-k_\text{max})/4$, where
$\text{tr}K=\Sigma_i k_i$ and $k_\text{max}=\max\{k_1,k_2,k_3\}$.
For quantum states with the X structure (of course, including those
of the form of Eq. (9)) one can obtain the eigenvalues of $K$
analytically as
\begin{eqnarray}
 k_{1,2}&=&4(|\rho_{14}|\pm|\rho_{23}|)^2,\nonumber\\
 k_{3}&=&2\sum_{n=1}^4
 \rho_{nn}^2-4(\rho_{11}\rho_{33}+\rho_{22}\rho_{44}).
\end{eqnarray}

The $k_i$ are in arbitrary order and cannot be ordered by magnitude
unless the parameter values are known, thus $k_\text{max}$ may be
changed with the variation of the parameters involved, which causes
the discontinuities of the GMQD. Two exemplified plots for certain
specified system parameters are presented in Fig. 4, from which one
can see that for fixed $\alpha^2=0.9$, the GMQD is obtained when
$k_\text{max}=k_3$ in the short and long time regions, while in the
middle time region, it is obtained when $k_\text{max}=k_1$. For
fixed $\gamma t=0.35$, we have $k_\text{max}=k_3$ for small or large
values of $\alpha^2$, and $k_\text{max}=k_1$ for middle values of
$\alpha^2$.

In Ref. \cite{Dakic} Daki\'{c} {\it et al.} showed that
$k_\text{max}$ can be written as
$k_\text{max}=\max_{|\vec{e}=1|}\vec{e}^T K\vec{e}$, and the sudden
change in decay rates of GMQD corresponds to the sudden change of
the optimized
$\vec{e}=(\sin\theta\cos\phi,\sin\theta\sin\phi,\cos\theta)^T$ in
$\mathbb{R}^3$. For the two-qubit case, the closest zero-discord
state $\chi=\Sigma_{k=1}^{2} p_k\Pi_k\otimes\rho_k$ is obtained when
$\Pi_1=(\mathbb{I}+\vec{e}\cdot\vec{\sigma})/2$ and
$\Pi_2=\mathbb{I}-\Pi_1$, with $\vec{e}$ being the eigenvector of
$K$ with the largest eigenvalue. The eigenvectors correspond to the
three eigenvalues of Eq. (12) are
\begin{eqnarray}
 |\psi_{1,2}\rangle&=&a_{1,2}\left(1,\frac{\text{Re}(\rho_{14}\rho_{23})
                      \mp|\rho_{14}\rho_{23}|}{\text{Im}(\rho_{14}\rho_{23})},0\right)^T,\nonumber\\
 |\psi_{3}\rangle&=&(0,0,1)^T,
\end{eqnarray}
where $a_{1,2}$ are normalization constants. $\vec{e}$ assumes
$|\psi_1\rangle$ or $|\psi_3\rangle$ with the variation of the
involved system parameters. If $\vec{e}=|\psi_1\rangle$ then we have
$\cos\theta=0$ or equivalently, $\theta=(n+0.5)\pi$ with
$n\in\mathbb{Z}$. If $\vec{e}=|\psi_3\rangle$, however, we have
$\cos\theta=\pm1$ or $\theta=n\pi$ with $n\in\mathbb{Z}$. The
optimal $\phi$ can also be derived exactly, and here we do not care
about it for the sudden change behaviors of GMQD have already been
reflected by the discontinuity of the optimal $\theta$.

Notice that although both the optimal unit vectors $\vec{n}$ for the
QD and $\vec{e}$ for the GMQD are related to the projective
operators, they are not the same for general case, which is
reflected in the disagreement of their individual sudden changes in
the decay rates.
\begin{figure}
\centering
\resizebox{0.4\textwidth}{!}{%
\includegraphics{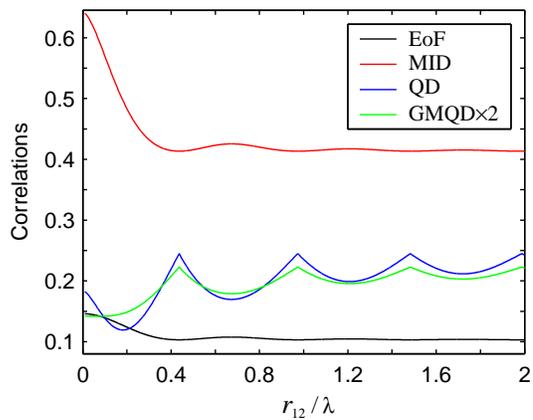}}
\caption{(Color online) Quantum correlations versus $r_{12}/\lambda$
for the initial state $|\Psi\rangle$ with $\alpha^2=0.8$ and $\gamma
t=0.35$.} \label{fig:5}
\end{figure}

In Fig. 5 we show the influence of $r_{12}/\lambda$ on correlation
dynamics for the initial state $|\Psi\rangle$ with $\alpha^2=0.8$
and $\gamma t=0.35$. It is seen that all the correlation measures
behave as damped oscillations with increasing $r_{12}/\lambda$, and
the dependence of QD and GMQD on $r_{12}/\lambda$ are more sensitive
than those of EoF and MID. Also one can note that there exists
states $\rho_1$ and $\rho_2$  such that
$\text{QD}(\rho_1)>\text{QD}(\rho_2)$,
$\text{GMQD}(\rho_1)>\text{GMQD}(\rho_2)$ and
$\text{MID}(\rho_1)<\text{MID}(\rho_2)$. Thus we see again that the
states have different orderings induced by different measures of
nonclassical correlations.
\begin{figure}
\centering
\resizebox{0.45\textwidth}{!}{%
\includegraphics{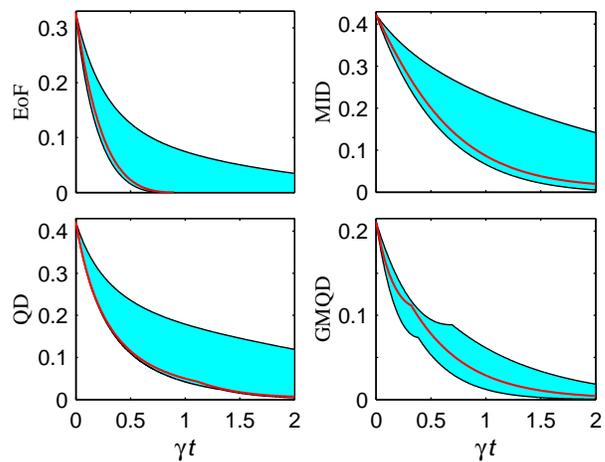}}
\caption{(Color online) Quantum correlations versus $\gamma t$ for
the initial state
$\rho^{\Psi}(0)=p|\Psi\rangle\langle\Psi|+(1-p)\mathbb{I}/4$ with
$\alpha^2=1/2$, $p=0.65$ and $r_{12}=0.6737\lambda$ (red curves).
The cyan regions show the extent to which the correlations can be
adjusted by performing operations $\{\mathbb{I}\times V\}$ to the
second qubit.} \label{fig:6}
\end{figure}

Finally, we show that the robustness of quantum correlations may be
enhanced by performing local unitary operations to the initial state
$\rho(0)$. As we know, the correlation measures (EoF, MID, QD and
GMQD) are locally unitary invariant; that is, $\rho(0)$ and
$(U\otimes V)\rho(0) (U^\dag\otimes V^\dag)$ have completely the
same values of correlations for any unitary operators $U$ and $V$
acting on parties $a$ and $b$. But this does not mean that their
robustness against decoherence are the same. In fact, for any
initial state $\rho(0)$, there exist an optimal local unitary
transformation $U_{\rm opt}\otimes V_{\rm opt}$ which can enhance
the robustness of the correlations to a certain greatest extent.
However, in general, the optimal $U_{\rm opt}\otimes V_{\rm opt}$ is
determined by the decoherence mechanism as well as the explicit form
of $\rho(0)$, and it is difficult to obtain an analytical expression
of it.

For simplicity, we consider the initial state as a Werner-like state
$\rho^{\Psi}(0)=p|\Psi\rangle\langle\Psi|+(1-p)\mathbb{I}/4$. And we
restrict ourselves to the complete set of locally unitary
transformations $\{\mathbb{I}\times V\}$ (a subset of $\{U\otimes
V\}$). Density matrix $\rho^{\Psi}(t)$ for $\rho^{\Psi}(0)$ can be
obtained analytically \cite{Ficek2}, and the dependence of EoF, MID,
QD as well as GMQD on the elements of $\rho^{\Psi}(t)$ are the same
as those for the initial state $|\Psi\rangle$. But the parameter
appeared in Eq. (11) changes to
$\delta=2|\rho_{14}||\rho_{23}|\cos(2\phi-\omega_1-\omega_2)$, where
$\cos\omega_{1,2}=\text{Re}(\rho_{14,23})/|\rho_{14,23}|$ and
$\sin\omega_{1,2}=-\text{Im}(\rho_{14,23})/|\rho_{14,23}|$. For
general $\rho(0)\in\{(\mathbb{I}\times
V)\rho^{\Psi}(0)(\mathbb{I}\times V^\dag)\}$, however, the density
matrix $\rho(t)$ as well as the correlation measures can only be
obtained numerically since the X structure of it may be destroyed
under the transformation $\{\mathbb{I}\times V\}$. By expressing $V$
in terms of the Pauli operators as
$V=\sin\alpha(\sigma^1\cos\gamma+\sigma^2\sin\gamma)+\cos\alpha
(\sigma^3\cos\beta-{\text i}\sigma^0\sin\beta)$ with
$\alpha\in[0,\pi/2]$ and $\beta, \gamma\in[0,2\pi]$, we perform
numerical calculations with $p=0.65$, $\alpha^2=1/2$,
$r_{12}=0.6737\lambda$ and show the results in Fig. 6. It is shown
that the extent of different correlation measures at finite time
$t>0$ can be adjusted among values bounded by the cyan regions by
performing operations $\{\mathbb{I}\times V\}$ to the initial state.
Particularly, the sudden death phenomenon of entanglement may be
avoided. Thus we see that different locally unitary equivalent
states show different robustness against decoherence induced by the
spontaneous emission, and this provides a way for long-time
preservation of quantum correlations in this system.

\section{Conclusions}
In conclusion, we evaluated systematically the quantum correlations
quantified by the EoF, MID, QD and GMQD by exactly solving a model
with the two qubits interacting with a multimode radiation field. We
showed that the dynamics for these quantities may be rather
different. Particularly, while entanglement experiences sudden
death, MID, QD and GMQD persist for an infinite time, i.e., they are
more robust than that of entanglement against decoherence, which is
important for QIP tasks which deos not based on entanglement.
Moreover, the dissipative process of spontaneous emission may lead
to a generation of states manifesting the relativity of different
correlation measures. In general, different measures of nonclassical
correlations are incomparable since their behaviors may be
qualitatively different.

We also demonstrated the disagreement between QD and GMQD on
reflecting the sudden change behaviors of nonclassical correlations
of a given quantum state. The origin of this phenomenon is the
different definitions of discord based on different measures of
distance between two states. While the QD is based on the von
Neumann entropy, the GMQD is based on the Hilbert-Schmidt norm, and
it is this difference that yields different optimal measurement
operators.

Since it is generally accepted that quantum correlations is crucial
for QIP, it is important to maintain them as long as possible. Here
we showed that although local unitary operations cannot change
extents of quantum correlations of a state, it can greatly enhance
the robustness of a system against decoherence. This provides a
possible way for long-time preservation of the quantum correlations
by local unitary operations.
\\

\begin{center}
\textbf{ACKNOWLEDGMENTS}
\end{center}

This work was supported by NSFC (10974247), ``973'' program
(2010CB922904), NSF of Shaanxi Province (2010JM1011, 2009JQ8006),
and the Scientific Research Program of Education Department of
Shaanxi Provincial Government (2010JK843).

\newcommand{\PRL}{Phys. Rev. Lett. }
\newcommand{\PRA}{Phys. Rev. A }
\newcommand{\JPA}{J. Phys. A }
%

%

\end{document}